# Couple stress theories: Theoretical underpinnings and practical aspects from a new energy perspective

Ali R. Hadjesfandiari, Gary F. Dargush

Department of Mechanical and Aerospace Engineering
University at Buffalo, The State University of New York, Buffalo, NY 14260 USA

ah@buffalo.edu, gdargush@buffalo.edu

November 25, 2016

**Abstract**

In this paper, we examine theoretical and practical aspects of several versions of couple stress theory. This includes indeterminate Mindlin-Tiersten-Koiter couple stress theory (MTK-CST), indeterminate symmetric modified couple stress theory (M-CST) and determinate skew-symmetric consistent couple stress theory (C-CST). We observe that MTK-CST and M-CST not only suffer from inconsistencies, these theories also cannot describe properly several elementary deformations, such as pure torsion of a circular bar and pure bending of a plate. By using an energy method, we also demonstrate another aspect of the inconsistency of the indeterminate MTK-CST and M-CST for elastic solids. This is achieved by deriving the governing equilibrium equations for elastic bodies in MTK-CST, M-CST and C-CST. This development shows that the direct minimization of the total potential energy for MTK-CST and M-CST violates the divergence free compatibility condition of the rotation vector field, that is, $\omega_{i,i} = 0$. On the other hand, the direct minimization of the total potential energy for C-CST satisfies this compatibility condition automatically. This result demonstrates another aspect of the inner consistency of C-CST.

**Keywords:** Couple stress theories; Size-dependent mechanics; Indeterminacies; Curvature tensors; Variational methods; Lagrange multipliers

# 1. Introduction

Mindlin and Tiersten (1962) and Koiter (1964) developed the initial version of the couple stress theory, based on the Cosserat continuum theory (Cosserat and Cosserat, 1909), in which the deformation is completely specified by the continuous displacement field $u_i$. Therefore, the kinematical quantities, such as the rotation vector $\omega_i$, and measures of deformation, such as the strain tensor $e_{ij}$ and rotation gradient tensor $\omega_{i,j}$ are derived from this displacement field. As a result, Mindlin-Tiersten-Koiter couple stress theory (MTK-CST) is based on the rigid body portion of motion of infinitesimal elements of matter at each point of the continuum (Hadjesfandiari and Dargush, 2015a). In these important developments, Mindlin, Tiersten and Koiter correctly established that five geometrical and five mechanical boundary conditions can be specified on a smooth surface. In MTK-CST, the couple-stress tensor is energetically conjugate to the gradient of rotation vector $\omega_{j,i}$, which is taken as the curvature tensor. Although these researchers made a significant step forward for continuum mechanics, the final MTK-CST suffers from some serious inconsistencies and difficulties with the underlying formulations (Hadjesfandiari and Dargush, 2015b). The three main inconsistencies of MTK-CST for isotropic linear elastic materials are:

1. The indeterminacy in the spherical part of the couple-stress tensor and, as a result, in the skew-symmetric part of the force-stress tensor;

2. The inconsistency in boundary conditions, since the normal component of the couple-traction vector appears in the formulation;

3. The appearance of two couple-stress elastic coefficients $\eta$ and $\eta'$ for linear elastic isotropic material, although only one of these elastic coefficients, $\eta$, appears in the final governing equations when written in terms of displacements.

The appearance of an arbitrary spherical couple stress component in the couple-stress tensor is the result of the deviatoric or trace free $\omega_{i,i} = 0$ character of the bend-twist tensor $\omega_{i,j}$ in this theory. The disturbing character of this spherical component is that it does not create any deformation in the body, which means its effect is equivalent to a zero loading condition. This spherical part of the couple-stress tensor remains indeterminate when the rotation field $\omega_i$ is prescribed on the whole boundary, and cannot be simply ignored. It cannot also be determined in a consistent systematic way in most cases, when the normal couple-traction vector is specified on the whole or some part of the boundary surface. Eringen (1968) realized this inconsistency as a major mathematical problem in the original MTK-CST, which he afterwards called indeterminate couple stress theory.

A symmetric couple stress theory still suffers from the same inconsistencies and difficulties with the underlying formulation. In this theory, the symmetric part of the gradient of rotation vector field $\omega_{i,j}$ is taken as the curvature tensor. However, careful examination shows that this symmetric tensor is actually a torsion tensor (Hadjesfandiari and Dargush, 2011, 2015a). Consequently, the couple-stresses in this theory create a combination of torsion and anticlastic deformation with negative Gaussian curvature for surface elements of the continuum. As in the original MTK-CST, the spherical part of the couple-stress tensor also remains indeterminate in this theory and cannot be determined in a consistent systematic way. This theory originates from the work of Yang et al. (2002), which is commonly called the modified couple stress theory (M-CST). In their development, Yang et al. (2002) consider an extra artificial equilibrium equation for the moment of couples, in addition to the two vectorial force and moment equilibrium equations of the classical continuum. Application of this unsubstantiated equilibrium equation, apparently leads to a symmetric couple-stress tensor. It seems the main motivation for Yang et al. (2002) in their development has been to reduce the number of couple-stress material parameters for linear isotropic elastic material from two coefficients $\eta$ and $\eta'$ in the original Mindlin-Tiersten-Koiter theory to only one coefficient $\eta$.

Recently, Hadjesfandiari and Dargush (2011, 2015a) have developed the consistent couple stress theory (C-CST), which resolves all inconsistencies in the original MTK-CST. The triumph of this development is discovering the subtle skew-symmetric character of the couple-stress tensor, which reduces the number of independent stress components to nine. As a result, the curvature tensor in C-CST is the skew-symmetric part of the gradient of rotation vector field $\omega_{i,j}$. The fundamental step in this development is satisfying the requirement that the normal component of the couple-traction vector must vanish on the boundary surface in a systematic way. This is what Mindlin, Tiersten and Koiter missed in their important developments, although they correctly established the consistent boundary conditions. It is interesting to notice that the skew-symmetric character of the couple-stress tensor immediately resolves the indeterminacy problem by establishing that there is no spherical component. As a result, the couple-stress tensor is determinate in the skew-symmetric C-CST. It is important to notice that the skew-symmetric couple-stresses create ellipsoidal cap-like deformation with positive Gaussian curvature for surface elements of the continuum. Although the reason for developing C-CST has not been to reduce the number of couple-stress material coefficients, it turns out that for linear isotropic elastic material this theory requires only one couple-stress material parameter $\eta$.

Because of inconsistencies, it is nearly impossible to find a true solution for many problems in MTK-CST and M-CST that satisfies all boundary conditions. This can be observed in very elementary practical problems. For example, there is no consistent solution for pure torsion of a circular bar in these theories. We notice that the inconsistent approximate solutions for pure torsion in these theories predict significant size effect, which does not agree with experiments (Hadjesfandiari and Dargush, 2016). MTK-CST and M-CST also cannot describe pure bending of a plate properly (Hadjesfandiari, et al. 2016). Particularly, M-CST predicts no couple-stresses and no size effect for the pure bending of the plate into a spherical shell. On the other hand, C-CST predicts consistent results for pure torsion of a circular bar and pure bending of a plate.

In this paper, we take a fresh view by examining the validity of MTK-CST, M-CST and C-CST from an energy perspective. Energy methods provide convenient and alternative means for formulating the governing equations of continuum solid mechanics. These methods not only give new insight into the formulations, but can also be used for formulating effective methods to obtain approximate solutions. Therefore, we concentrate on elastic bodies and derive the governing equilibrium equations and boundary conditions for the indeterminate MTK-CST and M-CST and determinate skew-symmetric C-CST. We minimize the total potential energy functional corresponding to these theories subject to the compatibility condition of the rotation vector field $\omega_{i,i} = 0$. Therefore, we impose this compatibility constraint by using the Lagrange multiplier method. As we shall see, for MTK-CST and M-CST, the Lagrange multiplier is the indeterminate spherical part of the couple-stress tensor. This means that the variational method results in a spherical component for the couple-stress tensor in the original Mindlin-Tiersten-Koiter couple stress theory (MTK-CST) and symmetric modified couple stress theory (M-CST). Therefore, the couple-stress tensor does not become trace free as one might conclude from an incorrect direct unconstrained minimization. This result clearly demonstrates that the direct unconstrained minimization of the total potential energy for MTK-CST and M-CST violates the divergence free compatibility condition of rotation vector field, such that $\omega_{i,i} = 0$. Thus, another aspect of the inconsistency of indeterminate MTK-CST and M-CST is revealed based on the energy method. On the other hand, for C-CST, the corresponding Lagrange multiplier vanishes, which demonstrates the consistency of the skew-symmetric character of the couple-stress tensor in this theory. Therefore, the direct unconstrained minimization of C-CST satisfies automatically the compatibility condition of rotation vector field $\omega_{i,i} = 0$. This result shows another inner consistency of C-CST.

The remainder of the paper is organized as follows. In Section 2, we present a brief review of the couple stress theory for linear isotropic elastic materials. This includes presenting MTK-CST, M-CST and C-CST, and examining their consistency from a theoretical and practical view. The main new results are then provided in Section 3, where we derive the governing equations for elastic bodies by minimizing the total mechanical potential energy for MTK-CST, M-CST and C-CST, subject to $\omega_{i,i} = 0$. This includes examining the consistency of MTK-CST, M-CST and C-CST from an energy perspective. Finally, Section 4 contains a summary and some general conclusions.

## 2. Couple stress theory

Consider a material continuum occupying a volume $V$ bounded by a surface $S$ with outer unit normal $n_i$, as shown in Fig. 1.

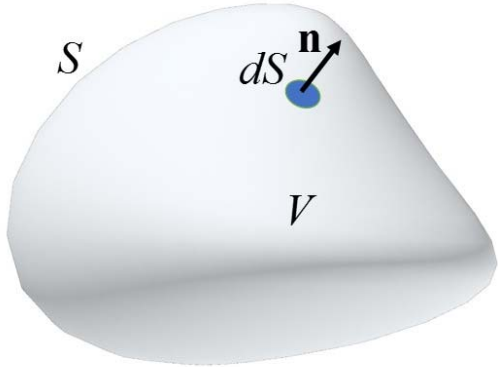

**Fig. 1.** The body configuration.

In couple stress theory, the interaction in the body is represented by true (polar) force-stress $\sigma_{ij}$ and pseudo (axial) couple-stress $\mu_{ij}$ tensors. The components of the force-stress $\sigma_{ij}$ and couple-stress $\mu_{ij}$ tensors in this theory are shown in Fig. 2.

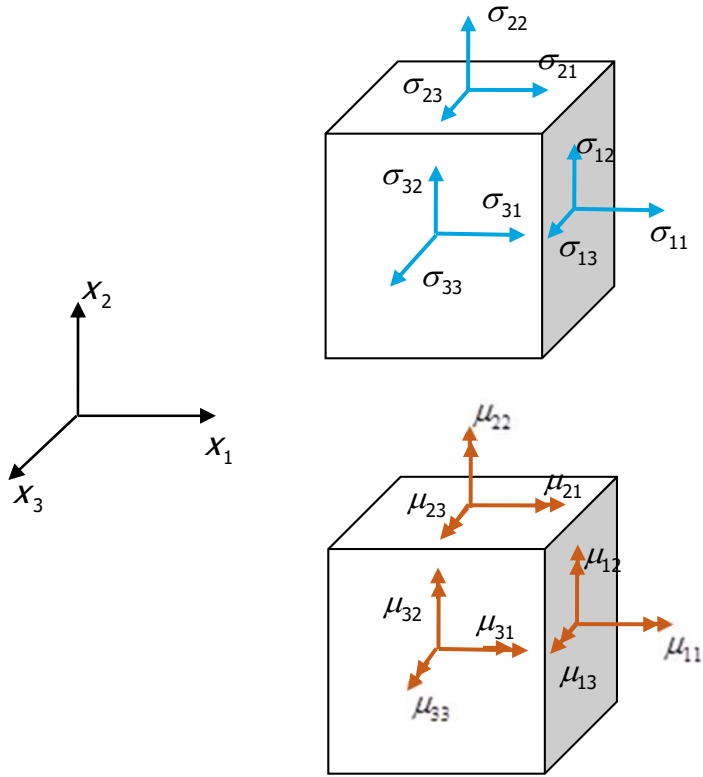

**Fig. 2.** General components of force- and couple-stress tensors in the couple stress theory.

As in the Cosserat continuum theory (Cosserat and Cosserat,1909), the force and moment balance governing equations for an infinitesimal element of matter under quasistatic conditions are written, respectively, as:

$$\sigma_{ji,j} + F_i = 0 \tag{1}$$

$$\mu_{ji,j} + \varepsilon_{ijk}\sigma_{jk} = 0 \tag{2}$$

where $F_i$ is the specified body-force density and $\varepsilon_{ijk}$ is the Levi-Civita alternating symbol. The force-traction vector $t_i$ and couple-traction vector $m_i$ at a point on surface element $dS$ with unit normal vector $n_i$ are given by

$$t_i = \sigma_{ji} n_j \tag{3}$$

$$m_i = \mu_{ji} n_j \tag{4}$$

The force-stress tensor is generally non-symmetric and can be decomposed as

$$\sigma_{ji} = \sigma_{(ji)} + \sigma_{[ji]} \tag{5}$$

where $\sigma_{(ji)}$ and $\sigma_{[ji]}$ are the symmetric and skew-symmetric parts, respectively. The angular equilibrium equation (2) gives the skew-symmetric part of the force-stress tensor as

$$\sigma_{[ji]} = \frac{1}{2}\varepsilon_{ijk}\mu_{lk,l} \tag{6}$$

Thus, for the total force-stress tensor, we have

$$\sigma_{ji} = \sigma_{(ji)} + \frac{1}{2}\varepsilon_{ijk}\mu_{lk,l} \tag{7}$$

As a result, the linear equation of equilibrium (1) reduces to

$$\left[\sigma_{(ji)} + \frac{1}{2}\varepsilon_{ijk}\mu_{lk,l}\right]_{,j} + F_i = 0 \tag{8}$$

In infinitesimal deformation theory, the displacement vector field $u_i$ is sufficiently small that the infinitesimal strain and rotation tensors are defined as

$$e_{ij} = u_{(i,j)} = \frac{1}{2}\left(u_{i,j} + u_{j,i}\right) \tag{9}$$

$$\omega_{ij} = u_{[i,j]} = \frac{1}{2}\left(u_{i,j} - u_{j,i}\right) \tag{10}$$

respectively. Since the true (polar) tensor $\omega_{ij}$ is skew-symmetrical, one can introduce its corresponding dual axial (pseudo) rotation vector as

$$\omega_i = \frac{1}{2}\varepsilon_{ijk}\omega_{kj} = \frac{1}{2}\varepsilon_{ijk}u_{k,j} \tag{11}$$

We notice that the definition (11) requires

$$\omega_{i,i} = 0 \tag{12}$$

which is the compatibility equation for the pseudo rotation vector. This condition constrains the form of a given rotation vector $\omega_i$.

It should be noticed that the rigid body portion of motion associated with infinitesimal elements (or rigid triads) at each point of the continuum is represented by the displacement vector $u_i$ and the rotation vector $\omega_i$. As demonstrated in Hadjesfandiari and Dargush (2015a), the suitable measures or metrics of deformation are defined based on the relative rigid body motion of triads at adjacent points of the continuum. This means the measures of deformation are defined based on $u_{i,j}$ and $\omega_{i,j}$. We should emphasize here that in classical infinitesimal theory only the symmetric part of $u_{i,j}$ defines the deformation. Should we expect that the entire tensor $\omega_{i,j}$ contribute as a measure of deformation in couple stress theory?

Since the tensor $\omega_{i,j}$ represents a combination of bending and torsional deformation of the material at each point, it can be properly called the bend-twist tensor (deWit, 1973). The infinitesimal pseudo (axial) torsion and mean curvature tensors (Hadjesfandiari and Dargush, 2011) are defined, respectively, as

$$\chi_{ij} = \omega_{(i,j)} = \frac{1}{2}\left(\omega_{i,j} + \omega_{j,i}\right) \tag{13}$$

$$\kappa_{ij} = \omega_{[i,j]} = \frac{1}{2}\left(\omega_{i,j} - \omega_{j,i}\right) \tag{14}$$

Since the mean curvature tensor $\kappa_{ij}$ is also skew-symmetrical, we can define its corresponding dual polar (true) mean curvature vector as

$$\kappa_i = \frac{1}{2}\varepsilon_{ijk}\kappa_{kj} \tag{15}$$

This can also be expressed as

$$\kappa_i = \frac{1}{2}\omega_{ji,j} = \frac{1}{4}\left(u_{j,ji} - \nabla^2 u_i\right) \tag{16}$$

Mindlin and Tiersten (1962) and Koiter (1964) have shown that the displacement field $u_i$ specified on a smooth part of the boundary surface $S$, specifies the normal component of the rotation $\omega^{(nn)} = \omega_i n_i$. Accordingly, they have demonstrated that material in a consistent couple stress theory does not support independent distributions of normal surface couple (or twisting) traction $m^{(nn)} = m_i n_i$. This means

$$m^{(nn)} = m_i n_i = \mu_{ji} n_i n_j = 0 \quad \text{on } S \tag{17}$$

From a mathematical point of view, these results show that we can specify either the displacement vector $u_i$ or the force-traction vector $t_i$, and the tangential component of the rotation vector $\omega_i$ or the tangent couple-traction vector $m_i$. In the other words, for three-dimensional boundary value problems, the number of kinematical and mechanical boundary conditions are each five. However, Mindlin, Tiersten and Koiter did not realize the fundamental implication of equation (17) as a constraint on the form of the couple-stress tensor $\mu_{ij}$ (Hadjesfandiari and Dargush, 2011, 2015a), which postponed the definition of a consistent couple stress theory for half a century.

For boundary conditions in general couple stress theory, one may specify displacements $u_i$ or force-tractions $t_i$

$$u_i = \bar{u}_i \quad \text{on } S_u \tag{18a}$$

$$t_i = \bar{t}_i \quad \text{on } S_t \tag{18b}$$

and tangential rotation $\omega_i$ or bending couple-traction $m_i$

$$\omega_i = \bar{\omega}_i \quad \text{on } S_\omega \tag{19a}$$

$$m_i = \bar{m}_i \quad \text{on } S_m \tag{19b}$$

Here $S_u$ and $S_\omega$ are the portions of the surface at which the essential boundary values for the displacement vector $u_i$ and the rotation $\omega_i$ are prescribed, respectively. Furthermore, $S_t$ and $S_m$ are the portions of the surface at which the force-traction vector $t_i$ and the couple-traction $m_i$ are specified, respectively. In order to construct a well-posed boundary value problem, we must have

$$S_u \cup S_t = S, \quad S_u \cap S_t = \varnothing \tag{20a}$$

$$S_\omega \cup S_m = S, \quad S_\omega \cap S_m = \varnothing \tag{20b}$$

Now we present specific aspects of the original indeterminate couple stress theory (MTK-CST), modified symmetric couple stress theory (M-CST) and consistent skew-symmetric couple stress theory (C-CST).

### 2.1. Original couple stress theory (MTK-CST)

In the original Mindlin-Tiersten-Koiter couple stress theory (MTK-CST), there is no constraint on the couple-stress tensor. As a result, there are 15 independent stress components. This includes six components of $\sigma_{(ij)}$ and nine components of $\mu_{ij}$.

The corresponding measure of deformation conjugate to the couple-stress tensor is the bend-twist tensor

$$k_{ij} = \omega_{j,i} \tag{21}$$

where

$$k_{ii} = \omega_{i,i} = 0 \tag{22}$$

For linear isotropic elastic material, the constitutive relations are

$$\sigma_{(ij)} = \lambda e_{kk}\delta_{ij} + 2\mu e_{ij} \tag{23}$$

$$\begin{aligned}\mu_{ij} &= Q\delta_{ij} + 4\eta k_{ij} + 4\eta' k_{ji} \\ &= Q\delta_{ij} + 4\eta \omega_{j,i} + 4\eta' \omega_{i,j}\end{aligned} \tag{24}$$

Here $Q\delta_{ij}$ is the indeterminate spherical part of the couple-stress tensor, where $Q$ is a pseudo-scalar. We notice that the moduli $\lambda$ and $\mu$ are the Lamé coefficients for isotropic media. These two coefficients are related by

$$\lambda = 2\mu \frac{\nu}{1-2\nu} \tag{25}$$

where $\nu$ is the Poisson's ratio. The parameters $\eta$ and $\eta'$ are the couple stress material coefficients for isotropic media.

For this material, the elastic energy density takes the form

$$W = \frac{1}{2}\lambda(e_{kk})^2 + \mu e_{ij}e_{ij} + 2\eta k_{ij}k_{ji} + 2\eta' k_{ij}k_{ij} \tag{26}$$

where the positive-definite elastic energy condition requires

$$3\lambda + 2\mu > 0, \quad \mu > 0, \quad \eta > 0, \quad -\eta < \eta' < \eta \tag{27a-d}$$

We can define the ratios

$$\frac{\eta}{\mu} = l^2, \qquad \frac{\eta'}{\eta} = c \tag{28a,b}$$

where

$$-1 < c < 1 \tag{29}$$

Here $l$ defines a characteristic material length, which accounts for size-dependency in this theory. Therefore, the relation (24) can be written as

$$\mu_{ij} = Q\delta_{ij} + 4\mu l^2 \left( \omega_{j,i} + c\omega_{i,j} \right) \tag{30}$$

We notice that the indeterminacy of $Q$ then carries into the skew-symmetrical part of the force-stress tensor, such that

$$\sigma_{[ji]} = \frac{1}{2} \varepsilon_{ijk} \mu_{lk,l}$$
$$= \frac{1}{2} \varepsilon_{ijk} Q_{,k} + 2\mu l^2 \varepsilon_{ijk} \nabla^2 \omega_k \tag{31}$$

and the total force-stress tensor becomes

$$\sigma_{ji} = \frac{1}{2} \varepsilon_{ijk} Q_{,k} + \lambda e_{kk} \delta_{ij} + 2\mu e_{ij} + 2\mu l^2 \varepsilon_{ijk} \nabla^2 \omega_k \tag{32}$$

which also is indeterminate. Therefore, we obtain the linear equilibrium equation in terms of the displacement as

$$\left[ \lambda + \mu\left(1 + l^2 \nabla^2\right) \right] u_{k,ki} + \mu(1 - l^2 \nabla^2)\nabla^2 u_i + F_i = 0 \tag{33}$$

In indeterminate couple stress theory (MTK-CST), the couple-traction is

$$m_i = \left( Q\delta_{ij} + 4\eta \omega_{i,j} + 4\eta' \omega_{j,i} \right) n_j$$
$$= Qn_i + 4\left( \eta \omega_{i,j} + \eta' \omega_{j,i} \right) n_j \tag{34}$$

As a result, the normal surface couple (or twisting) traction $m^{(nn)}$ becomes

$$m^{(nn)} = m_i n_i$$
$$= Q + 4(\eta + \eta')\omega_{i,j} n_i n_j \quad (35)$$

However, this component is not necessarily zero as required in (17), even if we ignore the indeterminacy term $Q$. This contradiction obviously shows that the MTK-CST is inconsistent. To resolve this problem, we may apparently use the transformation proposed by Koiter (1964) that a distribution of normal surface twisting couple-traction $m^{(nn)}$ on the actual surface $S$ is replaced by an equivalent shear stress distribution and a line force system. However, this transformation is not consistent with the idea of a continuum mechanics theory (Hadjesfandiar and Dargush, 2015a). A consistent couple stress theory must satisfy this condition directly in its formulation, that is

$$m^{(nn)} = m_i n_i = 0 \quad \text{on } S \quad (36)$$

Therefore, all troubles in the indeterminate couple stress theory (MTK-CST) are the result of not satisfying this condition in a systematic way. Let us impose this constraint at the present stage and investigate its consequences on MTK-CST. We notice that by the fundamental continuum mechanics hypothesis (Hadjesfandiar and Dargush, 2015a), the normal surface twisting couple-traction $m^{(nn)}$ must not only vanish on the actual boundary surface $S$, but on the boundary surface $S_a$ of any arbitrary subdomain with volume $V_a$, as shown in Fig. 3, that is

$$m^{(nn)} = Q + 4(\eta + \eta')\omega_{i,j} n_i n_j = 0 \quad \text{on } S_a \quad (37)$$

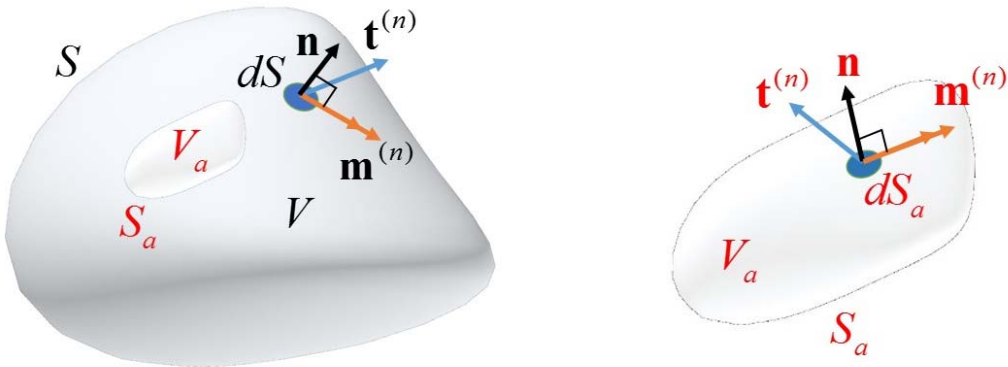

**Fig. 3.** The state of couple-traction $\mathbf{m}^{(n)}$ inside the body.

However, in this relation, $n_i$ is arbitrary at each point and $\omega_{i,j}$ can be any arbitrary deviatoric tensor; we may construct subdomains with any surface normal orientation at a point. Consequently, the condition (37) requires that

$$Q = 0 \quad \text{and} \quad \eta + \eta' = 0 \tag{38}$$

However, the condition (27d), rewritten equivalently as $0 < \eta + \eta' < 2\eta$, requires

$$Q = 0 \quad \text{and} \quad \eta = \eta' = 0 \tag{39}$$

which means that there are no couple-stresses in the body. Therefore, the consistency condition has reduced MTK-CST to the classical theory. We should mention that Koiter's loading transformation method is just an approximation method, which conceals the inconsistency of this continuum theory in an unreasonable manner. Therefore, finding solutions in MTK-CST that satisfy all boundary conditions consistently is practically impossible for many problems. For example, there is no consistent solution for pure torsion of a circular bar in this theory, and the approximate solution for pure torsion in MTK-CST predicts significant size effect, which does not agree with experiments (Hadjesfandiari and Dargush, 2016). In addition, MTK-CST cannot describe the pure bending of a plate properly (Hadjesfandiari, et al. 2016).

## 2.2. Modified couple stress theory (M-CST)

In this couple stress theory originally proposed by Yang et al. (2002), the pseudo couple-stress tensor is symmetrical, that is,

$$\mu_{ji} = \mu_{ij} \tag{40}$$

Therefore, there are 12 independent stress components in this theory. This includes six components of $\sigma_{(ij)}$ and six components of $\mu_{ij}$. We notice that in this theory, the normal couple-stress components $\{\mu_{11}, \mu_{22}, \mu_{33}\}$ on the plane element surfaces create torsion, and tangential components $\{\mu_{12}, \mu_{13}, \mu_{23}\}$ deform these plane elements to anticlastic surfaces with negative Gaussian curvature. Therefore, the symmetric couple stress tensor creates a combination of torsion and anticlastic deformation for plane element surfaces of a continuum.

The corresponding measure of deformation conjugate to the symmetric couple-stress tensor is the symmetric deviatoric torsion tensor $\chi_{ij}$

$$\chi_{ij} = \frac{1}{2}\left(\omega_{i,j} + \omega_{j,i}\right) \tag{41}$$

where

$$\chi_{ii} = \omega_{i,i} = 0 \tag{42}$$

For linear isotropic elastic material, the constitutive relations are

$$\sigma_{(ij)} = \lambda e_{kk}\delta_{ij} + 2\mu e_{ij} \tag{43}$$

$$\begin{aligned}\mu_{ij} &= Q\delta_{ij} + 8\mu l^2 \chi_{ij} \\ &= Q\delta_{ij} + 4\mu l^2 \left(\omega_{i,j} + \omega_{j,i}\right)\end{aligned} \tag{44}$$

We notice that the couple stress constant $\eta$ can be expressed as

$$\eta = \mu l^2 \tag{45}$$

The relations in the modified couple stress theory (M-CST) in original form (Yang et al., 2002) can be found by scaling $\eta \to \eta/4$ and $l \to l/2$ in the present equations. The elastic energy density for this material takes the form

$$W = \frac{1}{2}\lambda\left(e_{kk}\right)^2 + \mu e_{ij}e_{ij} + 4\eta \chi_{ij}\chi_{ij} \tag{46}$$

We notice that the indeterminacy of $Q$ in (44) then carries into the skew-symmetrical part of the force-stress tensor, such that

$$\sigma_{[ji]} = \frac{1}{2}\varepsilon_{ijk}\mu_{lk,l}$$
$$= \frac{1}{2}\varepsilon_{ijk}Q_{,k} + 2\mu l^2 \varepsilon_{ijk}\nabla^2 \omega_k \tag{47}$$

and the total force-stress tensor becomes

$$\sigma_{ji} = \frac{1}{2}\varepsilon_{ijk}Q_{,k} + \lambda e_{kk}\delta_{ij} + 2\mu e_{ij} + 2\mu l^2 \varepsilon_{ijk}\nabla^2 \omega_k \tag{48}$$

In this theory, for the linear equilibrium equation in terms of the displacement, we obtain exactly the same equation as (33) in MTK-CST, that is

$$\left[\lambda + \mu\left(1 + l^2 \nabla^2\right)\right] u_{k,ki} + \mu(1 - l^2 \nabla^2)\nabla^2 u_i + F_i = 0 \tag{49}$$

In modified couple stress theory (M-CST), the couple-traction is

$$m_i = \left(Q\delta_{ij} + 8\mu l^2 \chi_{ij}\right) n_j$$
$$= Qn_i + 4\mu l^2 \left(\omega_{i,j} + \omega_{j,i}\right) n_j \tag{50}$$

As a result, the normal surface couple (or twisting) traction $m^{(nn)}$ becomes

$$m^{(nn)} = m_i n_i$$
$$= Q + 8\mu l^2 \omega_{i,j} n_i n_j \tag{51}$$

We notice that in M-CST this component is not necessarily zero as required in (17), even if we ignore the indeterminacy term $Q$. This contradiction shows that the modified couple stress theory is also inconsistent. As mentioned above in Section 2.1, the Koiter transformation is not consistent with the idea of a continuum mechanics theory (Hadjesfandiar and Dargush, 2015a). A consistent couple stress theory must satisfy this condition directly in the formulation, that is

$$m^{(nn)} = m_i n_i = 0 \tag{52}$$

Therefore, all troubles in the M-CST, as in MTK-CST, are the result of not satisfying this zero normal couple traction condition in a systematic way. By imposing this constraint for the boundary surface $S_a$ of any arbitrary subdomain with volume $V_a$, as shown in Fig. 3, we obtain

$$m^{(nn)} = Q + 8\mu l^2 \omega_{i,j} n_i n_j = 0 \quad \text{on} \quad S_a \tag{53}$$

However, in this relation, $n_i$ is arbitrary at each point and $\omega_{i,j}$ can be any arbitrary deviatoric tensor, because we may construct subdomains with any surface normal orientation at a point. Consequently, the condition (53) requires that

$$Q = 0 \quad \text{and} \quad \eta = \mu l^2 = 0 \tag{54}$$

which means that there is no couple-stresses in the body. Therefore, the consistency condition has reduced the modified couple stress theory to the classical theory.

From a practical point of view, it is also generally impossible to satisfy all boundary conditions correctly in the solution of most problems using M-CST. This can be observed for the pure torsion of a circular bar, where it is impossible to obtain an exact continuum M-CST solution. Furthermore, the apparent approximate pure torsion solution predicts a significant size-effect, which contradicts recent experiments for pure torsion of micro-diameter copper wires (Hadjesfandiari and Dargush, 2016). In addition, M-CST cannot describe the pure bending of a plate properly (Hadjesfandiari, et al. 2016). Surprisingly, M-CST predicts no couple-stresses and no size effect for the pure bending of the plate into a spherical shell. These characteristics also make M-CST of questionable value to serve as a basis for size-dependent structural models, such as beams, plates and shells. Perhaps, this character of M-CST has not been fully understood, but in any case, it is unfortunate that this theory has been used so extensively in structural mechanics.

### 2.3. Consistent couple stress theory (C-CST)

In this couple stress theory, the pseudo couple-stress tensor is skew-symmetrical

$$\mu_{ji} = -\mu_{ij} \tag{55}$$

Therefore, there are nine independent stress components in this theory. This includes six components of $\sigma_{(ij)}$ and three components of $\mu_{ij}$. Since the couple-stress tensor is skew-symmetric, the couple-traction $m_i$ given by (4) becomes tangent to the surface. As a result, the couple-stress tensor $\mu_{ij}$ creates only bending couple-traction on any arbitrary surface in matter. The components of the force-stress $\sigma_{ij}$ and couple-stress $\mu_{ij}$ tensors in this theory are shown in Fig. 4.

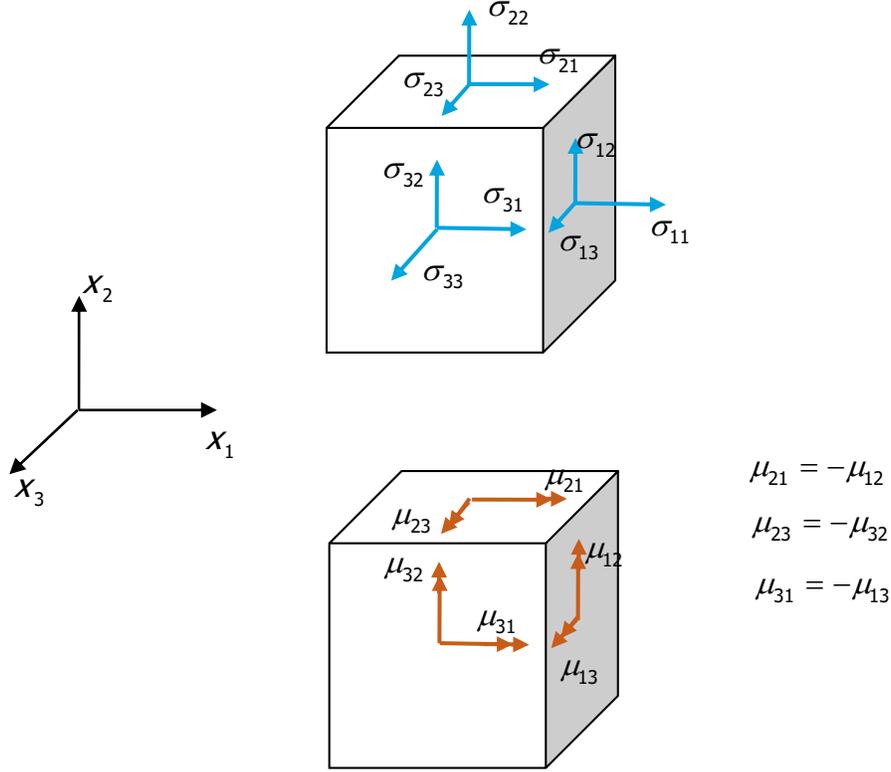

**Fig. 4.** Components of force- and couple-stress tensors in consistent couple stress theory.

We notice that in this theory the couple-stress components deform the plane element surfaces to ellipsoid cap-like surfaces with positive Gaussian curvature. It should be emphasized that the skew-symmetric character of the couple-stress tensor is a fundamental continuum mechanics property, which has nothing to do with any constitutive relation. As a consequence, this result is in no way limited to linear elastic materials or to isotropic response.

In C-CST, we can define the true (polar) couple-stress vector $\mu_i$ dual to the tensor $\mu_{ij}$ as

$$\mu_i = \frac{1}{2}\varepsilon_{ijk}\mu_{kj} \tag{56}$$

This relation can also be written in the form

$$\varepsilon_{ijk}\mu_k = \mu_{ji} \tag{57}$$

Consequently, the surface couple-traction vector $m_i$ reduces to

$$m_i = \mu_{ji}n_j = \varepsilon_{ijk}n_j\mu_k \tag{58}$$

which obviously is tangent to the surface.

The corresponding measure of deformation conjugate to the skew-symmetric couple-stress tensor $\mu_{ij}$ is the skew-symmetric mean curvature tensor $\kappa_{ij}$.

We notice that the skew-symmetric part of the force-stress tensor from equation (6) becomes

$$\sigma_{[ji]} = -\mu_{[i,j]} \tag{59}$$

Thus, for the total force-stress tensor, we have

$$\sigma_{ji} = \sigma_{(ji)} + \frac{1}{2}\varepsilon_{ijk}\mu_{lk,l} = \sigma_{(ji)} - \mu_{[i,j]} \tag{60}$$

As a result, the linear equation of equilibrium (1) reduces to

$$\left[\sigma_{(ji)} + \mu_{[j,i]}\right]_{,j} + F_i = 0 \tag{61}$$

For linear isotropic elastic material, the constitutive relations are

$$\sigma_{(ij)} = \lambda e_{kk}\delta_{ij} + 2\mu e_{ij} \tag{62}$$

$$\begin{aligned}\mu_{ij} &= -8\mu l^2 \kappa_{ij} \\ &= -4\mu l^2 \left(\omega_{i,j} - \omega_{j,i}\right)\end{aligned} \tag{63}$$

Here we have $l$ as the characteristic material length in the consistent couple stress theory, where the couple stress constant $\eta$ can be expressed as

$$\eta = \mu l^2 \tag{64}$$

The elastic energy density in this theory takes the form

$$W = \frac{1}{2}\lambda(e_{kk})^2 + \mu e_{ij}e_{ij} + 4\eta \kappa_{ij}\kappa_{ij} \tag{65}$$

which can also be written as

$$W = \frac{1}{2}\lambda(e_{kk})^2 + \mu e_{ij}e_{ij} + 8\mu l^2 \kappa_i \kappa_i \tag{66}$$

Then, by using (59), we obtain

$$\sigma_{[ji]} = 2\mu l^2 \varepsilon_{ijk} \nabla^2 \omega_k \tag{67}$$

for the skew-symmetric part of the force-stress tensor. Therefore, the total force-stress tensor becomes

$$\sigma_{ji} = \lambda e_{kk}\delta_{ij} + 2\mu e_{ij} + 2\mu l^2 \varepsilon_{ijk}\nabla^2 \omega_k \tag{68}$$

which is fully determinate. Interestingly, for the linear equilibrium equation for isotropic elastic materials in terms of the displacement, we obtain exactly the same equation as (33) and (49) in MTK-CST and M-CST, that is

$$\left[\lambda + \mu(1+l^2\nabla^2)\right]u_{k,ki} + \mu(1-l^2\nabla^2)\nabla^2 u_i + F_i = 0 \tag{69}$$

In the consistent couple stress theory (C-CST), the couple traction is

$$\begin{aligned} m_i &= \mu_{ji}n_j \\ &= -4\mu l^2 \left(\omega_{j,i} - \omega_{i,j}\right)n_j \end{aligned} \tag{70}$$

We notice that the normal surface couple-traction $m^{(nn)}$ vanishes, that is,

$$\begin{aligned} m^{(nn)} &= m_i n_i = \mu_{ji} n_j n_i \\ &= -4\mu l^2 \left(\omega_{j,i} - \omega_{i,j}\right)n_j n_i = 0 \end{aligned} \tag{71}$$

Therefore, there is no indeterminacy and no normal surface couple-traction $m^{(nn)}$. This is the main reason that this theory is consistent. As a result, problems within C-CST can be well-posed and boundary conditions can be satisfied precisely. For example, for the pure torsion of an elastic circular bar, the solution in C-CST reduces to that in classical theory, where there is no size effect. Interestingly, this prediction completely agrees with recent experiments for pure torsion of micro-diameter copper wires (Hadjesfandiari and Dargush, 2016). In addition, C-CST is the only couple stress theory, which describes the pure bending of a plate properly (Hadjesfandiari, et al. 2016).

It should be mentioned that discovering the skew-symmetric character of the pseudo couple-stress tensor $\mu_{ij}$ results in defining the skew-symmetric pseudo mean curvature tensor $\kappa_{ij}$. Since the couple-stress and mean curvature have true vectorial character, we can define the true couple-stress vector $\mu_i$ and mean curvature vector $\kappa_i$. Interestingly, this result gives the clue how to define the mean curvature vector in different space and higher dimensions. It turns out that this has a fundamental impact in understanding some physical phenomena. For example, Hadjesfandiari (2013) has developed the geometrical vortex theory of electromagnetism in four-dimensional space-time, where the electromagnetic four-vector potential and strength fields are the four-dimensional velocity and vorticity fields, respectively. This theory shows that the magnetic and the electric fields are the circular and hyperbolic vorticity-like fields, respectively. Therefore, the homogeneous Maxwell's equations are the necessary compatibility equations for the electromagnetic vorticity vectors, whereas the inhomogeneous Maxwell's equations govern the motion of these vorticities. Geometrically, the inhomogeneous Maxwell's equations are the relation for the mean curvature four-vector of the electromagnetic velocity field. It turns out that these equations simply show that the four-vector electric current density is proportional to the four-dimensional mean curvature of the four-vector potential field.

## 2.4. Discussion

In this section, we have presented different versions of couple stress theories (MTK-CST, M-CST and C-CST) and examined some of their theoretical and practical aspects. We have noticed that MTK-CST and M-CST not only suffer from different inconsistencies, such as indeterminacy and ill-posed boundary conditions, these theories also cannot describe properly several elementary practical problems, such as pure torsion of a circular bar and pure bending of a plate. On the other hand, C-CST is consistent with well-posed boundary conditions and can describe the pure torsion of a circular bar and pure bending of a plate properly.

Although the final governing equations for isotropic linear elastic materials in terms of the displacement vector are the same in these theories, the distribution of internal stresses and boundary conditions are different. Because of the indeterminacy, it is practically impossible to find solutions to problems in MTK-CST and M-CST, which satisfy all boundary conditions consistently. We notice that the inconsistent approximate solutions for pure torsion in these theories predict a significant size effect, which does not seem to agree with experiments (Hadjesfandiari and Dargush, 2016). The recent physical experiments by Song and Lu (2015), Liu et al. (2013) and Lu and Song (2011) entirely agree with the prediction of skew-symmetric consistent couple stress theory (C-CST) that there is no size effect in the elastic range for pure torsion of micro-diameter copper wires. Furthermore, MTK-CST and M-CST also cannot describe pure bending of a plate properly (Hadjesfandiari, et al. 2016). Particularly, M-CST predicts no couple-stresses and no size effect for the pure bending of the plate into a spherical shell.

Interestingly, Tang (1983) has applied MTK-CST based beam bending of Kao et al. (1979) and Tzung et al. (1981) to examine four-point bending and uniaxial tensile data of various size cylindrical and square specimens for three grades of graphite: H-327, H-451, and AGOT. The evaluations indicate that the data can be interpreted by linear couple stress theory (MTK-CST) with $\frac{\eta'}{\eta} \approx -0.85$ for H-451 graphite. Furthermore, the results are improved by considering a non-linear effect, which yields the new value $\frac{\eta'}{\eta} \approx -0.96$. As we can see, these result in hindsight show that the experimental data actually approach the consistent couple stress theory (C-CST) corresponding to $\frac{\eta'}{\eta} = -1$. On the other hand, one would need $\frac{\eta'}{\eta} \to +1$ for verification of M-CST, which clearly is not consistent with this experimental data for graphite.

In light of all of the above, we realize that the confusion in the development of couple stress theory for more than a half a century has given some status to MTK-CST and M-CST. Had Mindlin, Tiersten and Koiter discovered the skew-symmetric character of the couple-stress tensor in the 1960s, there would not have been the present confusion in couple stress theory. We should notice that the consistent couple stress theory (C-CST) systematically links efforts of Cosserats, Mindlin, Tiersten and Koiter and others in a span of a century.

Although the final linear equilibrium equations in terms of the displacement in all three different versions of couple stress theories (MTK-CST, M-CST and C-CST) are the same, the modified couple stress theory (M-CST) and consistent couple stress theory (C-CST) are not special cases of the original Mindlin-Tiersten-Koiter theory (MTK-CST). This is because the different curvature tensors in M-CST and C-CST are the symmetric and skew-symmetric parts of the bend-twist tensor $\omega_{i,j}$, respectively. It apparently seems that for isotropic linear elastic materials the formulations in M-CST and C-CST can be obtained by letting $\eta' = \eta$ and $\eta' = -\eta$, respectively, in the original MTK-CST. However, we must notice that these cases are excluded by condition (27d) ($-\eta < \eta' < \eta$) for the indeterminate MTK-CST. In addition, this peculiarity is only valid for isotropic material. There is no simple analogy for general anisotropic or non-linear cases.

Furthermore, a consistent theory such as C-CST cannot be considered as a special case of an inconsistent theory such as MTK-CST. However, we should remember that MTK-CST stands as a fundamental pillar in the development of the consistent couple stress theory (C-CST). This is obvious from the fact that elements of the C-CST are based on the original MTK-CST. Although the work of Mindlin, Tiersten and Koiter is fundamental in development of couple stress theory, they did not realize that satisfying the boundary condition $m^{(nn)} = 0$ in a systematic manner reveals the determinate skew-symmetric nature of the couple-stress tensor. To perceive better the status of MTK-CST, M-CST and C-CST in continuum mechanics, we can use the following simple illustrative analogy to classical continuum mechanics:

1. MTK-CST with $k_{ij} = \omega_{j,i}$ as the curvature tensor is analogous to a classical theory with $u_{i,j}$ as the strain tensor;

2. M-CST with $\chi_{ij}$ as the curvature tensor is analogous to a classical theory with $\omega_{ij}$ as the strain tensor;
3. C-CST with $\kappa_{ij}$ as the curvature tensor is analogous to the correct classical theory with $e_{ij}$ as the strain tensor.

However, we notice that the classical theories based on $u_{i,j}$ and $\omega_{ij}$ are not correct and are never considered as possible theories. It is also obvious that the consistent classical theory with $e_{ij}$ as the strain tensor is not considered as a special case of the general theory with $u_{i,j}$ as the strain tensor. Interestingly, it seems possible to solve some simple problems in the classical theories with $u_{i,j}$ and $\omega_{ij}$ as strain measures of deformation. For example, the simple tension or bending of a slender bar can be solved reasonably in all of these theories. However, these elementary solutions do not justify the theories based on $u_{i,j}$ and $\omega_{ij}$ as viable theories. As discussed, there is a similar situation in couple stress theories.

In recent papers (Neff et al., 2016; Ghiba et al., 2016; Madeo et al., 2016; Münch et al., 2015), the authors have claimed that the consistent skew-symmetric couple stress theory is not the only possible theory to represent the continuum consistently. They advocate that the other theories are also valid and can be arbitrarily used. For example, they suggest that the couple-stress tensor may be chosen symmetric and trace free (Münch et al., 2015). Although these papers use a labyrinth of mathematical formulae and are nearly impenetrable, the work is still limited to linear isotropic elasticity, rather than providing generality for continuum mechanics as a whole. Interestingly, these authors have also claimed discovering the correct traction boundary conditions in the indeterminate couple stress model (Neff et al., 2015). However, the newly defined boundary conditions in the indeterminate model are far too complicated and non-physical. It is also not known why the couple-stress tensor would still be indeterminate in a consistent model. These authors do not realize that the indeterminacy means there is still a trouble or inconsistency in this model, which is why some researchers gave up on MTK-CST and revived the idea of microrotation concept (Eringen, 1968).

Furthermore, it is incorrect to think that the energy method could justify ignoring the indeterminate spherical part of the couple-stress tensor. As will be demonstrated in the following section, the direct minimization of total potential energy for MTK-CST and M-CST violates the divergence free compatibility constraint $\omega_{i,i} = 0$, which has led Neff and colleagues to erroneous conclusions.

## 3. Variational method and its consequences

In this section, we derive the governing equations for an elastic body by using a variational method minimizing the total mechanical potential energy. We consider the original, modified and consistent couple stress theories (MTK-CST, M-CST and C-CST, respectively) and develop the energy method for the general non-linear anisotropic elastic case under infinitesimal deformation theory. It should be noticed that we ignore the aforementioned inconsistencies of MTK-CST and M-CST in our variational method in this section.

The total potential energy $\Pi$ for the elastic body is

$$\Pi = \int_V W dV - \int_V F_i u_i dV - \int_{S_t} \overline{t}_i u_i dS - \int_{S_m} \overline{m}_i \omega_i dS \tag{72}$$

where $S_t$ and $S_m$ are the portions of the surface on which $t_i$ and $m_i$ are prescribed, respectively. Here $W$ represents the general elastic energy density function, where

$$W = W(e_{ij}, k_{ij}) \quad \text{in the original couple stress theory (MTK-CST)} \tag{73a}$$

$$W = W(e_{ij}, \chi_{ij}) \quad \text{in modified couple stress theory (M-CST)} \tag{73b}$$

$$W = W(e_{ij}, \kappa_{ij}) \quad \text{in consistent couple stress theory (C-CST)} \tag{73c}$$

The equilibrium condition corresponds to the minimum of the total potential energy. It should be noticed that in a consistent theory, we do not need to impose the divergence free compatibility constraint

$$\omega_{i,i} = 0 \tag{74}$$

in the minimization. We expect that this constraint is satisfied in all steps of the minimization process. However, for assurance, we impose this constraint by using the Lagrange multiplier method and investigate if it vanishes in the final results. If the Lagrange multiplier persists within the formulation, then this indicates that the direct minimization of $\Pi$ cannot satisfy the divergence free compatibility constraint $\omega_{i,i} = 0$ at least in one of the steps of the process. Thus, there is an inconsistency in the corresponding couple stress theory formulation. Therefore, by using the Lagrange multiplier method to enforce the constraint (74), we define the Lagrangian functional

$$\tilde{\Pi} = \int_V W dV - \int_V F_i u_i dV - \int_{S_t} \overline{t}_i u_i dS - \int_{S_m} \overline{m}_i \omega_i dS + \int_V q \omega_{i,i} dV \tag{75}$$

where $q$ is the Lagrange multiplier that can be a function of space. The equilibrium condition corresponds to

$$\delta \tilde{\Pi} = 0 \tag{76}$$

where $\delta \tilde{\Pi}$ is the first variation of the functional $\tilde{\Pi}$. If $q = 0$ in the final results, then the constraint (74) is satisfied automatically in the minimization of $\Pi$ in (72). This would show that the corresponding couple stress theory is consistent. On the other hand, if $q$ does not vanish in the final result, then it indicates that the direct minimization of $\Pi$ in (72) violates the constraint (74). This means that there is some inconsistency in the corresponding couple stress theory. Interestingly, the physical meaning of the Lagrange multiplier $q$ will be revealed after obtaining the governing equations and the boundary conditions.

### 3.1. Variational method for original couple stress theory (MTK-CST)

For the original Mindlin-Tiersten-Koiter couple stress theory (MTK-CST), $W = W(e_{ij}, k_{ij})$, and we have

$$\tilde{\Pi}(u_i, \omega_i, e_{ij}, k_{ij}) = \int_V W dV - \int_V F_i u_i dV - \int_{S_t} \overline{t}_i u_i dS - \int_{S_m} \overline{m}_i \omega_i dS + \int_V q \omega_{i,i} dV \tag{77}$$

Therefore, the first variation of $\tilde{\Pi}$ is

$$\delta\tilde{\Pi} = \int_V \left( \frac{\partial W}{\partial e_{ij}} \delta e_{ij} + \frac{\partial W}{\partial k_{ij}} \delta k_{ij} \right) dV - \int_V F_i \delta u_i dS - \int_{S_t} \bar{t}_i \delta u_i dS - \int_{S_m} \bar{m}_i \delta\omega_i dS + \int_V q \delta\omega_{i,i} dV \tag{78}$$

where

$$\delta k_{ij} = \delta\omega_{i,j} \tag{79}$$

We should notice that the variation of $\delta k_{ij}$ must be consistent with the variation of constraint

$$\delta k_{ii} = \delta\omega_{i,i} = 0 \tag{80}$$

which has been imposed in (78) by using the Lagrange multiplier method.

Note that $\bar{t}_i^{(n)}$, $\bar{m}_i^{(n)}$ and $F_i$ are specified quantities, not subject to variation in (78). By considering the conditions $\delta u_i = 0$ on $S_u$ and $\delta\omega_i = 0$ on $S_\omega$, (78) can be written as

$$\delta\tilde{\Pi} = \int_V \left( \frac{\partial W}{\partial e_{ij}} \delta e_{ij} + \frac{\partial W}{\partial k_{ij}} \delta k_{ij} \right) dV - \int_V F_i \delta u_i dS - \int_S t_i \delta u_i dS - \int_S m_i \delta\omega_i dS + \int_V q \delta\omega_{i,i} dV \tag{81}$$

By some manipulation, we obtain

$$\delta\tilde{\Pi} = \int_V \frac{1}{2}\left( \frac{\partial W}{\partial e_{ij}} + \frac{\partial W}{\partial e_{ji}} \right) \delta u_{i,j} dV + \int_V \left( \frac{\partial W}{\partial k_{ij}} + q\delta_{ij} \right) \delta\omega_{i,j} dV$$
$$- \int_S F_i \delta u_i dS - \int_V t_i \delta u_i dS - \int_S m_i \delta\omega_i dS \tag{82}$$

By using integration by part on the second term, this becomes

$$\delta\tilde{\Pi} = \int_V \frac{1}{2}\left( \frac{\partial W}{\partial e_{ij}} + \frac{\partial W}{\partial e_{ji}} \right) \delta u_{i,j} dV$$
$$+ \int_V \left\{ \left[ \left( \frac{\partial W}{\partial k_{ij}} + q\delta_{ij} \right) \delta\omega_i \right]_{,j} - \left( \frac{\partial W}{\partial k_{ij}} + q\delta_{ij} \right)_{,j} \delta\omega_i \right\} dV \tag{83}$$
$$- \int_S F_i \delta u_i dS - \int_V t_i \delta u_i dS - \int_S m_i \delta\omega_i dS$$

At this stage, we notice that

$$\delta \omega_i = \frac{1}{2}\varepsilon_{ijk}\delta u_{k,j} \tag{84}$$

Therefore, the variation (83) can be written as

$$\begin{aligned}\delta\tilde{\Pi} =& \int_V \left[\frac{1}{2}\left(\frac{\partial W}{\partial e_{ij}}+\frac{\partial W}{\partial e_{ji}}\right)-\varepsilon_{mji}\frac{1}{2}\left(\frac{\partial W}{\partial k_{mn}}+q\delta_{mn}\right)_{,n}\right]\delta u_{i,j}\,dV \\ &+ \int_V \left[\left(\frac{\partial W}{\partial k_{ij}}+q\delta_{ij}\right)\delta\omega_i\right]_{,j}\,dV \\ &- \int_S F_i\delta u_i\,dS - \int_S t_i\delta u_i\,dS - \int_S m_i\delta\omega_i\,dS\end{aligned} \tag{85}$$

Again by using integration by parts, this becomes

$$\begin{aligned}\delta\tilde{\Pi} =& \int_V \left\{\left[\frac{1}{2}\left(\frac{\partial W}{\partial e_{ij}}+\frac{\partial W}{\partial e_{ji}}\right)-\varepsilon_{mji}\frac{1}{2}\left(\frac{\partial W}{\partial k_{mn}}+q\delta_{mn}\right)_{,n}\right]\delta u_i\right\}_{,j}\,dV \\ &- \int_V \left[\frac{1}{2}\left(\frac{\partial W}{\partial e_{ij}}+\frac{\partial W}{\partial e_{ji}}\right)-\varepsilon_{mji}\frac{1}{2}\left(\frac{\partial W}{\partial k_{mn}}+q\delta_{mn}\right)_{,n}\right]_{,j}\delta u_i\,dV \\ &+ \int_V \left[\left(\frac{\partial W}{\partial k_{ij}}+q\delta_{ij}\right)\delta\omega_i\right]_{,j}\,dV \\ &- \int_S F_i\delta u_i\,dS - \int_S t_i\delta u_i\,dS - \int_S m_i\delta\omega_i\,dS\end{aligned} \tag{86}$$

Now we apply the divergence theorem to the first and third terms in the volume integral in (86) and by noticing that

$$\frac{1}{2}\varepsilon_{mji}q_{,mj} = 0 \tag{87}$$

we obtain the relation

$$\delta \tilde{\Pi} =$$

$$-\int_V \left\{ \left[ \frac{1}{2}\left(\frac{\partial W}{\partial e_{ij}} + \frac{\partial W}{\partial e_{ji}}\right) - \varepsilon_{mji}\frac{1}{2}\left(\frac{\partial W}{\partial k_{mn}}\right)_{,n} \right]_{,j} + F_i \right\} \delta u_i dV$$

$$+\int_S \left\{ \left[ \frac{1}{2}\left(\frac{\partial W}{\partial e_{ij}} + \frac{\partial W}{\partial e_{ji}}\right) - \varepsilon_{mji}\frac{1}{2}\left(\frac{\partial W}{\partial k_{mn}} + q\delta_{mn}\right)_{,n} \right] n_j - t_i \right\} \delta u_i dS \qquad (88)$$

$$+\int_S \left[ \left(\frac{\partial W}{\partial k_{ij}} + q\delta_{ij}\right) n_j - m_i \right] \delta \omega_i dS$$

We should recall the conditions $\delta u_i = 0$ on $S_u$ and $\delta \omega_i = 0$ on $S_\omega$ to obtain the relation

$$\delta \tilde{\Pi} =$$

$$-\int_V \left\{ \left[ \frac{1}{2}\left(\frac{\partial W}{\partial e_{ij}} + \frac{\partial W}{\partial e_{ji}}\right) - \frac{1}{2}\varepsilon_{mji}\left(\frac{\partial W}{\partial k_{mn}}\right)_{,n} \right]_{,j} + F_i \right\} \delta u_i dV$$

$$+\int_{S_t} \left\langle \left\{ \frac{1}{2}\left(\frac{\partial W}{\partial e_{ij}} + \frac{\partial W}{\partial e_{ji}}\right) - \varepsilon_{mji}\frac{1}{2}\left[\left(\frac{\partial W}{\partial k_{mn}}\right)_{,n} + q_{,m}\right] \right\} n_j - \bar{t}_i \right\rangle \delta u_i dS \qquad (89)$$

$$+\int_{S_m} \left[ \left(\frac{\partial W}{\partial k_{ij}} + q\delta_{ij}\right) n_j - \bar{m}_i \right] \delta \omega_i dS$$

The variation $\delta u_i$ is arbitrary in the domain $V$ in (89). The variations of $\delta u_i$ and $\delta \omega_i$ are also arbitrary on the boundary surfaces $S_t$ and $S_m$, respectively. Therefore, the individual terms in the integrals must vanish separately and we have

$$\left[ \frac{1}{2}\left(\frac{\partial W}{\partial e_{ij}} + \frac{\partial W}{\partial e_{ji}}\right) - \frac{1}{2}\varepsilon_{mji}\left(\frac{\partial W}{\partial k_{mn}}\right)_{,n} \right]_{,j} + F_i = 0 \quad \text{in } V \qquad (90)$$

$$\bar{t}_i = \left\{ \frac{1}{2}\left(\frac{\partial W}{\partial e_{ij}} + \frac{\partial W}{\partial e_{ji}}\right) - \frac{1}{2}\varepsilon_{mji}\left[\left(\frac{\partial W}{\partial k_{mn}}\right)_{,n} + q_{,m}\right] \right\} n_j \quad \text{on } S_t \qquad (91)$$

$$\bar{m}_i = \left(\frac{\partial W}{\partial k_{ij}} + q\delta_{ij}\right) n_j \quad \text{on } S_m \qquad (92)$$

For linear isotropic material, where $W$ is given by (26), these equations reduce to

$$\left[\lambda + \mu\left(1 + l^2\nabla^2\right)\right]u_{k,ki} + \mu(1 - l^2\nabla^2)\nabla^2 u_i + F_i = 0 \quad \text{in } V \tag{93}$$

$$\bar{t}_i = \left(\lambda e_{kk}\delta_{ij} + 2\mu e_{ij} + 2\mu l^2 \varepsilon_{ijm}\nabla^2\omega_m + \frac{1}{2}\varepsilon_{ijm}q_{,m}\right)n_j \quad \text{on } S_t \tag{94}$$

$$\bar{m}_i = \left[4\mu l^2\left(\omega_{j,i} + c\omega_{i,j}\right) + q\delta_{ij}\right]n_j \quad \text{on } S_m \tag{95}$$

By comparing the governing equation (93) and boundary conditions (94) and (95) with their corresponding equations in section 2.1 for MTK-CST, we recognize the Lagrange multiplier $q$ as the spherical part of the couple-stress tensor $Q$, that is

$$q = Q \tag{96}$$

and obtain the general constitutive relations

$$\sigma_{(ji)} = \lambda e_{kk}\delta_{ij} + 2\mu e_{ij} \tag{97}$$

$$\mu_{ij} = Q\delta_{ij} + 4\mu l^2\left(\omega_{j,i} + c\omega_{i,j}\right) \tag{98}$$

$$\sigma_{[ji]} = \frac{1}{2}\varepsilon_{ijk}Q_{,k} + 2\mu l^2 \varepsilon_{ijk}\nabla^2\omega_k \tag{99}$$

As we can see the variational method shows that the couple-stress tensor is still indeterminate.

Here we have demonstrated that the constrained minimization of $\tilde{\Pi}$ produces the fundamental governing equations in the indeterminate couple stress theory (MTK-CST) for linear elastic isotropic materials. Interestingly, the Lagrange multiplier corresponding to the constraint (74) $\omega_{i,i} = 0$ is the indeterminate spherical part of the couple-stress tensor. This development shows that if we do not impose the constraint (74) in our variational method, the direct minimization process of $\Pi$ violates this constraint in the first step of variation in (78), where $\delta k_{ii}$ is not necessarily zero. Therefore, the incorrect direct minimization of $\Pi$ results in a trace free couple-stress tensor in the original couple stress theory. Remarkably, the fact that the constraint (74) is not satisfied automatically in the minimization of $\Pi$ in (72) demonstrates that there is some inconsistency in the Mindlin-Tiersten-Koiter couple stress theory (M-CST).

*3.2. Variational method for modified symmetric couple stress theory (M-CST)*

For the modified couple stress theory (M-CST) $W = W(e_{ij}, \chi_{ij})$, and we have

$$\tilde{\Pi}(u_i, \omega_i, e_{ij}, \chi_{ij}) = \int_V W dV - \int_V F_i u_i dV - \int_{S_t} \overline{t_i} u_i dS - \int_{S_m} \overline{m_i} \omega_i dS + \int_V q \omega_{i,i} dV \qquad (100)$$

Therefore, the first variation of $\tilde{\Pi}$ is

$$\delta\tilde{\Pi} = \int_V \left( \frac{\partial W}{\partial e_{ij}} \delta e_{ij} + \frac{\partial W}{\partial \chi_{ij}} \delta \chi_{ij} \right) dV - \int_V F_i \delta u_i dS - \int_{S_t} \overline{t_i} \delta u_i dS - \int_{S_m} \overline{m_i} \delta \omega_i dS + \int_V q \delta \omega_{i,i} dV \qquad (101)$$

where

$$\delta \chi_{ij} = \frac{1}{2} (\delta \omega_{i,j} + \delta \omega_{j,i}) \qquad (102)$$

We should notice that the variation of $\delta \chi_{ij}$ must be consistent with the variation of constraint

$$\delta \chi_{ii} = \delta \omega_{i,i} = 0 \qquad (103)$$

which has been imposed in (101) by using the Lagrange multiplier method.

Again by considering the conditions $\delta u_i = 0$ on $S_u$ and $\delta \omega_i = 0$ on $S_\omega$, we can write (101) as

$$\delta\tilde{\Pi} = \int_V \left( \frac{\partial W}{\partial e_{ij}} \delta e_{ij} + \frac{\partial W}{\partial \chi_{ij}} \delta \chi_{ij} \right) dV - \int_V F_i \delta u_i dS - \int_S t_i \delta u_i dS - \int_S m_i \delta \omega_i dS + \int_V q \delta \omega_{i,i} dV \qquad (104)$$

Therefore, this can be written as

$$\delta \tilde{\Pi} = \int_V \frac{1}{2}\left(\frac{\partial W}{\partial e_{ij}} + \frac{\partial W}{\partial e_{ji}}\right)\delta u_{i,j} dV + \int_V \left[\frac{1}{2}\left(\frac{\partial W}{\partial \chi_{ij}} + \frac{\partial W}{\partial \chi_{ji}}\right) + q\delta_{ij}\right]\delta \omega_{i,j} dV$$
$$-\int_S F_i \delta u_i dS - \int_V t_i \delta u_i dS - \int_S m_i \delta \omega_i dS \qquad (105)$$

By using integration by part on the second term, this becomes

$$\delta \tilde{\Pi} = \int_V \frac{1}{2}\left(\frac{\partial W}{\partial e_{ij}} + \frac{\partial W}{\partial e_{ji}}\right)\delta u_{i,j} dV$$
$$+\int_V \left\{\left\{\left[\frac{1}{2}\left(\frac{\partial W}{\partial \chi_{ij}} + \frac{\partial W}{\partial \chi_{ji}}\right) + q\delta_{ij}\right]\delta \omega_i\right\}_{,j} - \left[\frac{1}{2}\left(\frac{\partial W}{\partial \chi_{ij}} + \frac{\partial W}{\partial \chi_{ji}}\right) + q\delta_{ij}\right]_{,j}\delta \omega_i\right\} dV \qquad (106)$$
$$-\int_S F_i \delta u_i dS - \int_V t_i \delta u_i dS - \int_S m_i \delta \omega_i dS$$

where, we have

$$\delta \omega_i = \frac{1}{2}\varepsilon_{ijk}\delta u_{k,j} \qquad (107)$$

Therefore, the variation can be written as

$$\delta \tilde{\Pi} = \int_V \left\{\frac{1}{2}\left(\frac{\partial W}{\partial e_{ij}} + \frac{\partial W}{\partial e_{ji}}\right) - \varepsilon_{mji}\left[\frac{1}{4}\left(\frac{\partial W}{\partial \chi_{mn}} + \frac{\partial W}{\partial \chi_{nm}}\right) + \frac{1}{2}q\delta_{mn}\right]_{,n}\right\}\delta u_{i,j} dV$$
$$+\int_V \left\{\left[\frac{1}{2}\left(\frac{\partial W}{\partial \chi_{ij}} + \frac{\partial W}{\partial \chi_{ji}}\right) + q\delta_{ij}\right]\delta \omega_i\right\}_{,j} dV \qquad (108)$$
$$-\int_S F_i \delta u_i dS - \int_S t_i \delta u_i dS - \int_S m_i \delta \omega_i dS$$

Again by using integration by parts, this becomes

$$\delta \tilde{\Pi} = \int_V \left\{ \left[ \frac{1}{2} \left( \frac{\partial W}{\partial e_{ij}} + \frac{\partial W}{\partial e_{ji}} \right) - \varepsilon_{mji} \left[ \frac{1}{4} \left( \frac{\partial W}{\partial \chi_{mn}} + \frac{\partial W}{\partial \chi_{nm}} \right)_{,n} + \frac{1}{2} q_{,m} \right] \right] \delta u_i \right\}_{,j} dV$$

$$- \int_V \left\{ \frac{1}{2} \left( \frac{\partial W}{\partial e_{ij}} + \frac{\partial W}{\partial e_{ji}} \right) - \varepsilon_{mji} \left[ \frac{1}{4} \left( \frac{\partial W}{\partial \chi_{mn}} + \frac{\partial W}{\partial \chi_{nm}} \right) + \frac{1}{2} q \delta_{mn} \right]_{,n} \right\}_{,j} \delta u_i \, dV \quad (109)$$

$$+ \int_V \left\{ \left[ \frac{1}{2} \left( \frac{\partial W}{\partial \chi_{ij}} + \frac{\partial W}{\partial \chi_{ji}} \right) + q \delta_{ij} \right] \delta \omega_i \right\}_{,j} dV$$

$$- \int_S F_i \delta u_i \, dS - \int_S t_i \delta u_i \, dS - \int_S m_i \delta \omega_i \, dS$$

Now we apply the divergence theorem to the first and third terms in the volume integral in (109) and by noticing that

$$\frac{1}{2} \varepsilon_{mji} q_{,mj} = 0 \quad (110)$$

we obtain the relation

$$\delta \tilde{\Pi} =$$

$$- \int_V \left\{ \left[ \frac{1}{2} \left( \frac{\partial W}{\partial e_{ij}} + \frac{\partial W}{\partial e_{ji}} \right) - \frac{1}{4} \varepsilon_{mji} \left( \frac{\partial W}{\partial \chi_{mn}} + \frac{\partial W}{\partial \chi_{nm}} \right)_{,n} \right]_{,j} + F_i \right\} \delta u_i \, dV$$

$$+ \int_S \left\{ \left\{ \frac{1}{2} \left( \frac{\partial W}{\partial e_{ij}} + \frac{\partial W}{\partial e_{ji}} \right) - \varepsilon_{mji} \left[ \frac{1}{4} \left( \frac{\partial W}{\partial \chi_{mn}} + \frac{\partial W}{\partial \chi_{nm}} \right) + \frac{1}{2} q \delta_{mn} \right]_{,n} \right\} n_j - t_i \right\} \delta u_i \, dS \quad (111)$$

$$+ \int_S \left\{ \left[ \frac{1}{2} \left( \frac{\partial W}{\partial \chi_{ij}} + \frac{\partial W}{\partial \chi_{ji}} \right) + q \delta_{ij} \right] n_j - m_i \right\} \delta \omega_i \, dS$$

By applying the conditions $\delta u_i = 0$ on $S_u$ and $\delta \omega_i = 0$ on $S_\omega$, we obtain the relation

$$\delta \tilde{\Pi} =$$

$$-\int_V \left\{ \left[ \frac{1}{2}\left(\frac{\partial W}{\partial e_{ij}} + \frac{\partial W}{\partial e_{ji}}\right) - \frac{1}{4}\varepsilon_{mji}\left(\frac{\partial W}{\partial \chi_{mn}} + \frac{\partial W}{\partial \chi_{nm}}\right)_{,n} \right]_{,j} + F_i \right\} \delta u_i dV$$

$$+\int_{S_t} \left\{ \left[ \frac{1}{2}\left(\frac{\partial W}{\partial e_{ij}} + \frac{\partial W}{\partial e_{ji}}\right) - \varepsilon_{mji}\left[\frac{1}{4}\left(\frac{\partial W}{\partial \chi_{mn}} + \frac{\partial W}{\partial \chi_{nm}}\right)_{,n} + \frac{1}{2}q_{,m}\right] \right] n_j - \bar{t}_i \right\} \delta u_i dS \quad (112)$$

$$+\int_{S_m} \left\{ \left[ \frac{1}{2}\left(\frac{\partial W}{\partial \chi_{ij}} + \frac{\partial W}{\partial \chi_{ji}}\right) + \xi\delta_{ij} \right] n_j - \bar{m}_i \right\} \delta \omega_i dS$$

We notice that the variation $\delta u_i$ is arbitrary in the domain $V$ in (112); and the variations of $\delta u_i$ and $\delta \omega_i$ are also arbitrary on the boundary surfaces $S_t$ and $S_m$, respectively. Therefore, the individual terms in the integrals vanish separately and we have

$$\left[ \frac{1}{2}\left(\frac{\partial W}{\partial e_{ij}} + \frac{\partial W}{\partial e_{ji}}\right) - \frac{1}{4}\varepsilon_{mji}\left(\frac{\partial W}{\partial \chi_{mn}} + \frac{\partial W}{\partial \chi_{nm}}\right)_{,n} \right]_{,j} + F_i = 0 \quad \text{in} \quad V \quad (113)$$

$$\bar{t}_i = \left\{ \frac{1}{2}\left(\frac{\partial W}{\partial e_{ij}} + \frac{\partial W}{\partial e_{ji}}\right) - \varepsilon_{mji}\left[\frac{1}{4}\left(\frac{\partial W}{\partial \chi_{mn}} + \frac{\partial W}{\partial \chi_{nm}}\right)_{,n} + \frac{1}{2}q_{,m}\right] \right\} n_j \quad \text{on} \quad S_t \quad (114)$$

$$\bar{m}_i = \left[ \frac{1}{2}\left(\frac{\partial W}{\partial \chi_{ij}} + \frac{\partial W}{\partial \chi_{ji}}\right) + q\delta_{ij} \right] n_j \quad \text{on} \quad S_m \quad (115)$$

For linear isotropic material, where $W$ is given by (46), these equations reduce to

$$\left[\lambda + \mu\left(1 + l^2\nabla^2\right)\right]u_{k,ki} + \mu(1 - l^2\nabla^2)\nabla^2 u_i + F_i = 0 \quad \text{in} \quad V \quad (116)$$

$$\bar{t}_i = \left(\lambda e_{kk}\delta_{ij} + 2\mu e_{ij} + 2\mu l^2 \varepsilon_{ijm}\nabla^2 \omega_m + \frac{1}{2}\varepsilon_{ijm}q_{,m}\right)n_j \quad \text{on} \quad S_t \quad (117)$$

$$\bar{m}_i = \left(8\eta\chi_{ij} + q\delta_{ij}\right)n_j \quad \text{on} \quad S_m \quad (118)$$

By comparing the governing equation (116) and boundary conditions (117) and (118) with their corresponding equation in section 2.2 for M-CST, we recognize the Lagrange multiplier $q$ as the spherical part of the couple-stress tensor $Q$, that is

$$q = Q \tag{119}$$

and obtain the general constitutive relations

$$\sigma_{(ji)} = \lambda e_{kk} \delta_{ij} + 2\mu e_{ij} \tag{120}$$

$$\begin{aligned}\mu_{ji} &= Q\delta_{ij} + 8\mu l^2 \chi_{ji} \\ &= Q\delta_{ij} + 4\mu l^2 \left(\omega_{i,j} + \omega_{j,i}\right)\end{aligned} \tag{121}$$

$$\sigma_{[ji]} = \frac{1}{2}\varepsilon_{ijk}Q_{,k} + 2\mu l^2 \varepsilon_{ijk}\nabla^2 \omega_k \tag{122}$$

Therefore, we have demonstrated that the constrained minimization of $\tilde{\Pi}$ produces the fundamental governing equations in the indeterminate modified couple stress theory (M-CST) for the linear elastic isotropic materials. We notice that for this case the Lagrange multiplier corresponding to the constraint (74) ($\omega_{i,i} = 0$) is also the indeterminate spherical part of the couple stress tensor. This development shows that if we do not impose the constraint (74) in our variational method, the direct minimization process of $\Pi$ violates this constraint in the first step of variation in (101), where $\delta\chi_{ii}$ is not necessarily zero. Therefore, the incorrectly developed direct minimization of $\Pi$ results in a trace free couple-stress tensor in the modified symmetric couple stress theory. Since the constraint $\omega_{i,i} = 0$ is not satisfied automatically in the minimization of $\Pi$ in (72), one should realize that there is some inconsistency in the modified couple stress theory (M-CST).

### 3.3. Variational method for consistent couple stress theory (C-CST)

One might expect that considering the compatibility constraint (74) ($\omega_{i,i} = 0$) is not necessary in minimization of $\Pi$ for the skew-symmetric consistent couple stress theory (C-CST). We will see if the final result confirms this speculation.

For this case, the energy density is $W = W(e_{ij}, \kappa_{ij})$. Thus, we have

$$\tilde{\Pi}(u_i, \omega_i, e_{ij}, \kappa_{ij}) = \int_V W dV - \int_V F_i u_i dV - \int_{S_t} \bar{t}_i u_i dS - \int_{S_m} \bar{m}_i \omega_i dS + \int_V q \omega_{i,i} dV \tag{123}$$

The first variation of $\tilde{\Pi}$ is

$$\delta\tilde{\Pi} = \int_V \left( \frac{\partial W}{\partial e_{ij}} \delta e_{ij} + \frac{\partial W}{\partial \kappa_{ij}} \delta \kappa_{ij} \right) dV - \int_V F_i \delta u_i dS - \int_{S_t} \bar{t}_i \delta u_i dS - \int_{S_m} \bar{m}_i \delta \omega_i dS + \int_V q \delta \omega_{i,i} dV \tag{124}$$

where

$$\delta \kappa_{ij} = \frac{1}{2}(\delta \omega_{i,j} - \delta \omega_{j,i}) \tag{125}$$

This relation shows that we always have

$$\delta \kappa_{ii} = 0 \tag{126}$$

which does not depend on the variation of the constraint

$$\delta \omega_{i,i} = 0 \tag{127}$$

imposed by using the Lagrange multiplier method in (124). This character seems to imply that we do not need to impose the divergence free compatibility constraint (123) in the minimization process from beginning for this case. However, we go forward and demonstrate this interesting character by deriving the governing equilibrium equations and the corresponding boundary conditions.

By considering the conditions $\delta u_i = 0$ on $S_u$ and $\delta \omega_i = 0$ on $S_\omega$, (124) can be written as

$$\delta\tilde{\Pi} = \int_V \left( \frac{\partial W}{\partial e_{ij}} \delta e_{ij} + \frac{\partial W}{\partial \kappa_{ij}} \delta \kappa_{ij} \right) dV - \int_V F_i \delta u_i dS - \int_S t_i \delta u_i dS - \int_S m_i \delta \omega_i dS + \int_V q \delta \omega_{i,i} dV \tag{128}$$

By a similar method as before, we obtain the relation

$$\delta \tilde{\Pi} =$$

$$-\int_V \left\{ \left[ \frac{1}{2}\left(\frac{\partial W}{\partial e_{ij}}+\frac{\partial W}{\partial e_{ji}}\right) - \frac{1}{4}\varepsilon_{mji}\left(\frac{\partial W}{\partial \kappa_{mn}}-\frac{\partial W}{\partial \kappa_{nm}}\right)_{,n}\right]_{,j} + F_i \right\} \delta u_i dV$$

$$+\int_{S_t} \left\{ \left\{ \frac{1}{2}\left(\frac{\partial W}{\partial e_{ij}}+\frac{\partial W}{\partial e_{ji}}\right) - \varepsilon_{mji}\left[\frac{1}{4}\left(\frac{\partial W}{\partial \kappa_{mn}}-\frac{\partial W}{\partial \kappa_{nm}}\right) + \frac{1}{2}q\delta_{mn}\right]_{,n} \right\} n_j - \bar{t}_i \right\} \delta u_i dS \quad (129)$$

$$+\int_{S_m} \left\{ \left[ \frac{1}{2}\left(\frac{\partial W}{\partial \kappa_{ij}}-\frac{\partial W}{\partial \kappa_{ji}}\right) + q\delta_{ij} \right] n_j - \bar{m}_i \right\} \delta \omega_i dS$$

We notice that in (129) the variation $\delta u_i$ is arbitrary in the domain $V$, while the variations of $\delta u_i$ and $\delta \omega_i$ are also arbitrary on the boundary surfaces $S_t$ and $S_m$, respectively. Therefore, the individual terms in the integrals vanish separately and we have

$$\left[ \frac{1}{2}\left(\frac{\partial W}{\partial e_{ij}}+\frac{\partial W}{\partial e_{ji}}\right) - \frac{1}{4}\varepsilon_{mji}\left(\frac{\partial W}{\partial \kappa_{mn}}-\frac{\partial W}{\partial \kappa_{nm}}\right)_{,n}\right]_{,j} + F_i = 0 \quad \text{in } V \quad (130)$$

$$\bar{t}_i = \left\{ \frac{1}{2}\left(\frac{\partial W}{\partial e_{ij}}+\frac{\partial W}{\partial e_{ji}}\right) - \varepsilon_{mji}\left[\frac{1}{4}\left(\frac{\partial W}{\partial \kappa_{mn}}-\frac{\partial W}{\partial \kappa_{nm}}\right)_{,n} + \frac{1}{2}q_{,m}\right] \right\} n_j \quad \text{on } S_t \quad (131)$$

$$\bar{m}_i = \left[ \frac{1}{2}\left(\frac{\partial W}{\partial \kappa_{ij}}-\frac{\partial W}{\partial \kappa_{ji}}\right) + q\delta_{ij} \right] n_j \quad \text{on } S_m \quad (132)$$

However, we notice that the condition

$$m^{(nn)} = \bar{m}_i n_i = 0 \quad \text{on } S_m \quad (133)$$

requires that

$$\bar{m}_i n_i = \left[ \frac{1}{2}\left(\frac{\partial W}{\partial \kappa_{ij}}-\frac{\partial W}{\partial \kappa_{ji}}\right) + q\delta_{ij} \right] n_j n_i$$
$$= \frac{1}{2}\left(\frac{\partial W}{\partial \kappa_{ij}}-\frac{\partial W}{\partial \kappa_{ji}}\right) n_i n_j + q n_i n_i = 0 \quad \text{on } S_m \quad (134)$$

Since the expression $\left(\dfrac{\partial W}{\partial \kappa_{ij}} - \dfrac{\partial W}{\partial \kappa_{ji}}\right)$ is skew symmetric, the relation (134) shows that the Lagrange multiplier disappears, that is

$$q = 0 \tag{135}$$

Therefore, the boundary conditions (131) and (132) reduce to

$$\bar{t}_i = \left\{\dfrac{1}{2}\left(\dfrac{\partial W}{\partial e_{ij}} + \dfrac{\partial W}{\partial e_{ji}}\right) - \varepsilon_{mji}\left[\dfrac{1}{4}\left(\dfrac{\partial W}{\partial \kappa_{mn}} - \dfrac{\partial W}{\partial \kappa_{nm}}\right)_{,n}\right]\right\}n_j \quad \text{on} \ S_t \tag{136}$$

$$\bar{m}_i = \dfrac{1}{2}\left(\dfrac{\partial W}{\partial \kappa_{ij}} - \dfrac{\partial W}{\partial \kappa_{ji}}\right)n_j \quad \text{on} \ S_m \tag{137}$$

For linear isotropic material, where $W$ is given by (65), these equations reduce to

$$\left[\lambda + \mu\left(1 + l^2\nabla^2\right)\right]u_{k,ki} + \mu(1 - l^2\nabla^2)\nabla^2 u_i + F_i = 0 \quad \text{in} \ V \tag{138}$$

$$\bar{t}_i = \left(\lambda e_{kk}\delta_{ij} + 2\mu e_{ij} + 2\mu l^2 \varepsilon_{ijm}\nabla^2\omega_m\right)n_j \quad \text{on} \ S_t \tag{139}$$

$$\bar{m}_i = 8\eta\kappa_{ij}n_j \quad \text{on} \ S_m \tag{140}$$

By comparing the governing equation (138) and boundary conditions (139) and (140) with their corresponding equations in section 2.3 for C-CST, we obtain the general constitutive relations

$$\sigma_{(ji)} = \lambda e_{kk}\delta_{ij} + 2\mu e_{ij} \tag{141}$$

$$\mu_{ji} = 8\mu l^2 \kappa_{ij} \tag{142}$$

$$\sigma_{[ji]} = 2\mu l^2 \varepsilon_{ijk}\nabla^2 \omega_k \tag{143}$$

Therefore, the total force-stress tensor becomes

$$\sigma_{ji} = \lambda e_{kk}\delta_{ij} + 2\mu e_{ij} + 2\mu l^2 \varepsilon_{ijk}\nabla^2 \omega_k \tag{144}$$

As expected, the result $q=0$ shows that we do not need to impose the compatibility condition constraint $\omega_{i,i}=0$ in the minimization process of the total potential energy in the consistent skew-symmetric couple stress theory (C-CST) from the very beginning. Therefore, the variational methods used in the formulations developed in Darrall et al. (2014, 2015) and Salter and Richardson (2014) are mathematically consistent.

We can see that the direct minimization process of $\Pi$ can produce the fundamental governing equations in the consistent couple stress theory (C-CST) for linear elastic isotropic materials without violating the divergence free constraint $\omega_{i,i}=0$.

### 3.4. Discussion

By using an energy method in this section, we have demonstrated another aspect of the inconsistency of the indeterminate MTK-CST and M-CST for elastic solids. We have shown that the direct unconstrained minimization of the total potential energy $\Pi$ for these theories violates the divergence free compatibility condition $\omega_{i,i}=0$. On the other hand, the direct unconstrained minimization of total potential energy $\Pi$ for C-CST satisfies this compatibility condition automatically. This result once again demonstrates the inner consistency of C-CST.

In their variational formulation for modified couple stress theory (M-CST), Park and Gao (2008) ignored the indeterminacy of the couple-stress tensor completely. As a result, they considered the unconstrained minimization of total potential energy $\Pi$ for M-CST, which violates the divergence free compatibility constraint $\omega_{i,i}=0$. Forgetting to impose the divergence free constraint $\omega_{i,i}=0$ in their variational method, Neff and his colleagues also mistakenly concluded that a variational method results in a trace free couple-stress tensor in M-CST (Neff et al., 2016; Ghiba et al., 2016 Münch et al., 2015). How can the couple-stress tensor be trace free in M-CST, when the unconstrained minimization of $\Pi$ violates the divergence free compatibility constraint $\omega_{i,i}=0$?

Neff et al. (2016) have also argued that the indeterminacy of the spherical part of the couple-stress tensor is analogous to the behavior of an incompressible material under pressure. For an incompressible material, the incompressibility condition is

$$u_{i,i} = 0 \tag{145}$$

Because we can apply any pressure to an incompressible solid without changing its shape, the stress cannot be uniquely determined from the strains. For a linear isotropic incompressible elastic material in classical theory, the constitutive relation for force stresses are

$$\sigma_{ij} = -p\delta_{ij} + 2\mu e_{ij} \tag{146}$$

where $p$ specifies the negative of the spherical part of the force-stress tensor. As a result, the linear equilibrium equation in terms of the displacement in classical elasticity becomes

$$-p_{,i} + \mu \nabla^2 u_i + F_i = 0 \tag{147}$$

Interestingly, we notice that the elastic energy density for this incompressible case takes the form

$$W = \mu e_{ij} e_{ij} \tag{148}$$

As we know, the pressure stress distribution does not contribute to the internal work, because

$$\int_0^{e_{ij}} \sigma_{ji} de_{ij} = -p e_{ii} = -p u_{i,i} = 0 \tag{149}$$

Interestingly, we notice that the pressure $p$ in (146) becomes the Lagrange multiplier corresponding to the incompressibility condition (145) in a variational energy method (Spencer, 1980; Fosdick and Royer-Carfagni, 1999). However, it seems Neff and his colleagues did not realize the consequence of this analogy in their variational method in couple stress theory for the compatibility condition of the rotation vector field $\omega_{i,i} = 0$. As mentioned, they have forgotten to enforce this constraint in their variational energy method by using the Lagrange multiplier method for couple stress theory.

We should also notice that an incompressible material is a mathematical concept, and physically does not exist. The incompressibility condition (145) is just an artificial assumption to simplify cases of near-incompressibility. We notice that this relation is not the result of the general character of $u_{i,j}$ or definition of the strain tensor as

$$e_{ij} = \frac{1}{2}\left(u_{i,j} + u_{j,i}\right) \tag{150}$$

This means that the strain tensor $e_{ij}$ never becomes deviatoric in reality. Interestingly, for the linear isotropic elastic materials, the incompressibility condition (145) corresponds to Poisson's ratio $\nu = \frac{\lambda}{2(\lambda+\mu)} = \frac{1}{2}$, which is excluded based on the energy considerations (27a) that requires

$$-1 < \nu < 1/2 \tag{151}$$

On the other hand, the deviatoric characters, such that

$$k_{ii} = \chi_{ii} = 0 \tag{152}$$

are the result of the mathematical definition of $k_{ij}$ and $\chi_{ij}$ for any arbitrary material. As a result, the spherical part of the couple-stress tensor in MTK-CST and M-CST becomes indeterminate for all materials independent of the material behavior. It is this deviatoric character, which makes the tensors $k_{ij}$ and $\chi_{ij}$ unsuitable measures of bending deformation. We should notice that the incompressibility of material $u_{i,i} = 0$ is just an approximation for some special cases. However, the divergence free constraint of the rotation vector $\omega_{i,i} = 0$ is a mathematical constraint for all bodies based on definition, which has been the source of all kinds of confusion in the evolution of size-dependent continuum mechanics at least for half a century (Hadjesfandiari and Dargush, 2015b).

## 4. Conclusions

In this paper, we have examined some theoretical and practical aspects of the three primary couple stress theories, namely, MTK-CST, M-CST and C-CST. It has been shown that MTK-CST and M-CST not only suffer from different inconsistencies, such as indeterminacy and ill-posed boundary conditions, these theories also cannot describe elementary practical deformations, such as pure torsion of a circular bar and pure bending of a plate. On the other hand, C-CST is consistent with well-posed boundary conditions and can describe pure torsion of a circular bar and pure bending of a plate.

Furthermore, by using an energy method, we have also derived the governing equilibrium equations for elastic bodies in MTK-CST, M-CST and C-CST. This development shows that the direct minimization of the total potential energy $\Pi$ for MTK-CST and M-CST violates the divergence free compatibility condition of the rotation vector field, $\omega_{i,i} = 0$. This demonstrates another aspect of the inconsistency of the indeterminate MTK-CST and M-CST for elastic bodies based on the energy method. Therefore, from a mathematical standpoint, the total potential energy functional $\Pi$ corresponding to MTK-CST and M-CST must be minimized subject to the compatibility condition $\omega_{i,i} = 0$. Therefore, one way to impose this compatibility condition constraint is by using the Lagrange multiplier method. Through this approach, we find that in MTK-CST and M-CST, the Lagrange multiplier is the indeterminate spherical part of the couple-stress tensor. This means the variational method results in a spherical component for the couple-stress tensor in MTK-CST and M-CST. On the other hand, the direct unconstrained minimization of total potential energy $\Pi$ for C-CST satisfies this compatibility condition automatically, which once more demonstrates the inner consistency of C-CST.